\documentclass{PoS}
\usepackage[T1]{fontenc}
\usepackage[utf8]{inputenc}
\usepackage{amsmath}
\usepackage{amssymb}
\usepackage{amsfonts}
\usepackage{caption}

\newtheorem{theorem}{Theorem}[section]
\newenvironment{remark*}[1][Remark]{\begin{trivlist}
\item[\hskip \labelsep {\bfseries #1}]}{\end{trivlist}}

\renewcommand{\phi}{\varphi}

\newcommand{\fa}{\forall}
\newcommand{\bigtimes}{\mathop{\mbox{\LARGE$\boldsymbol\times$}}}
\newcommand{\vol}{\ensuremath{\mathrm{vol}}}
\renewcommand{\d}{\partial}
\newcommand{\Bp}[1][{}]{\ensuremath{\mathcal{B}^{#1}}}
\newcommand{\cn}[1][{}]{\ensuremath{\mathbb{C}^{#1}}}
\newcommand{\nn}[1][{}]{\ensuremath{\mathbb{N}^{#1}}}
\newcommand{\Df}[1][{}]{\ensuremath{\mathfrak{D}}}

\title{New polynomially exact integration rules on $U(N)$ and $SU(N)$}

\ShortTitle{New polynomially exact integration rules on $U(N)$ and $SU(N)$}

\author{\mbox{Andreas Ammon,$^d$} \mbox{\speaker{Tobias Hartung},$^a$} \mbox{Karl Jansen,$^b$} \mbox{Hernan Le\"{o}vey,$^c$} and \mbox{Julia Volmer$^b$}\\
        \llap{$^a$}King's College London, Department of Mathematics\\
        Strand, London WC2R 2LS, United Kingdom\\
        \llap{$^b$}NIC, DESY Zeuthen\\
        Platanenallee 6, D-15738 Zeuthen, Germany\\
        \llap{$^c$}Humboldt-University Berlin, Department of Mathematics\\
        Unter den Linden 6, D-10099 Berlin, Germany\\
        \llap{$^d$}IVU Traffic Technologies AG\\
        Bundesallee 88 , D-12161 Berlin, Germany\\
        E-mail: \email{andreas.ammon@desy.de}, \email{tobias.hartung@kcl.ac.uk}, \email{karl.jansen@desy.de}, \email{leovey@math.hu-berlin.de}, \email{julia.volmer@desy.de}}

\abstract{
In lattice Quantum Field Theory, we are often presented with integrals over polynomials of coefficients of matrices in $U(N)$ or $SU(N)$ with respect to the Haar measure. In some physical situations, e.g., in presence of a chemical potential, these integrals are numerically very difficult since their integrands are highly oscillatory which manifests itself in form of the sign problem. In these cases, Monte Carlo methods often fail to be adequate, rendering such computations practically impossible.

We propose a new class of integration rules on $U(N)$ and $SU(N)$ which are derived from polynomially exact rules on spheres. We will examine these quadrature rules and their efficiency at the example of a 0+1 dimensional QCD for a non-zero quark mass and chemical potential. In particular, we will demonstrate the failure of Monte Carlo methods in such applications and that we can obtain polynomially exact, arbitrary precision results using the new integration rules.
}

\FullConference{34th annual International Symposium on Lattice Field Theory\\
         24-30 July 2016\\
         University of Southampton, UK}

\begin{document}

\section{Introduction}
One of the greatest computational challenges in models of statistical and high energy physics is the sign problem \cite{troyer-wiese}. Hence, a myriad of techniques have been developed to address this problem, yet no general solution has been found to date \cite{gattringer-langfeld}. In high energy physics, for instance, the sign problem prevents a full understanding of the early universe and heavy ion collisions. Such questions require computations using lattice QCD with a non-zero chemical potential which are impossible using current techniques due to the appearance of large cancelation errors (cf., \cite{gattringer,sexty} for recent reviews). 

Hence, alternative methods need to be developed. For instance, we have proposed and tested Quasi Monte Carlo and iterated numerical integration techniques \cite{jansen-1,jansen-2}. Similarly, polynomially exact integration techniques \cite{bloch-1} and approaches using symmetrization \cite{bloch-2,bloch-3} have been studied. In particular, the symmetrization approach achieved stable results combining Monte Carlo with fairly small symmetry groups. The method to be proposed in the present work is based on \emph{complete symmetrization} and results in an arbitrarily precise evaluation of the integrals considered.

We will construct completely symmetric integration rules for $U(N)$ and $SU(N)$ (section~\ref{sec:quadrature}) which lead to polynomial exactness and test them using the $1$-dimensional QCD with chemical potential (cf., e.g., \cite{ravagli}) as an example (sections~\ref{sec:1dQCD}~and~\ref{sec:num-res}). Though the $1$-dimensional QCD is interesting in its own right as the strong coupling limit of QCD \cite{ilgenfritz-kripfganz}, we will use it as a benchmark only; especially since observables can be computed analytically allowing us to check the numerical results directly. In particular, we will compute the chiral condensate over a broad range of action parameters including values that are impossible (for all practical purposes) to address with standard Monte Carlo techniques.

\section{Construction of polynomially exact rules derived from spheres}\label{sec:quadrature}
The construction of polynomially exact rules on $SU(N)$ and $U(N)$ is based on Theorem~\ref{theorem:haar-push-forward}.

\begin{theorem}\label{theorem:haar-push-forward}
  Let $G_1$ and $G_2$ be topological groups, $\Bp(G_1)$ and $\Bp(G_2)$ their Borel $\sigma$-algebras, $\Phi:\ G_1\to G_2$ an isomorphism (i.e., a group isomorphism that is also a homeomorphism), $\gamma_1$ a (normalized) Haar measure on $G_1$, and $\gamma_2$ defined by $\fa A\in\Bp(G_2):\ \gamma_2(A):=\gamma_1([A]\Phi)$ where $[A]\Phi$ denotes the pre-set of $A$ under $\Phi$.

  Then, $\gamma_2$ is a (normalized) Haar measure.
\end{theorem}

Since $U(N)=SU(N)\rtimes U(1)$ holds where $\rtimes$ denotes the (outer) semi-direct product, $SU(2)\cong S^3$, and $SU(N)$ is a principal $SU(N-1)$-bundle over $S^{2N-1}$ \cite[equation (22.18)]{frankel}, we can write $U(N)$ and $SU(N)$ as products of spheres (up to a set of measure zero); more precisely, $U(N)\simeq\bigtimes_{j=1}^{N}S^{2j-1}\quad\text{and}\quad SU(N)\simeq\bigtimes_{j=2}^{N}S^{2j-1}$. Thus, given a homeomorphism $\Phi:\ \bigtimes_jS^{2j-1}\to G$, we can push the group structure of $G\in\{U(N),SU(N)\}$ to the product of spheres turning $\Phi$ into an isomorphism as in Theorem~\ref{theorem:haar-push-forward}. In other words,
\begin{align}
  \int_G\ f\ dh_G = \int_{\bigtimes_jS^{2j-1}}\ f\circ\Phi\ d\vol_{\bigtimes_jS^{2j-1}}
\end{align}
where $h_G$ is the normalized Haar measure on $G$ and $\vol_{\bigtimes_jS^{2j-1}}=\bigtimes_j\frac{\vol_{S^{2j-1}}}{\vol_{S^{2j-1}}(S^{2j-1})}$ is the product of the normalized volume measures on the spheres defined on the product $\sigma$-algebra $\bigotimes_j\Bp\left(S^{2j-1}\right)$.

The map $\Phi$ can be constructed inductively. Let $\Phi_2:\ \d B_{\cn[2]}\cong S^3\to SU(2);\ \begin{pmatrix}\alpha\\\beta\end{pmatrix}\mapsto\begin{pmatrix}\alpha&-\beta^*\\\beta&\alpha^*\end{pmatrix}$ where $\d B_X$ is the boundary of the unit ball in $X$ and $j_N:\ SU(N-1)\to SU(N);\ U\mapsto\begin{pmatrix}U&0\\0&1\end{pmatrix}$. Furthermore, let
\begin{align}\label{eq:R1}
  r_{j,N}:=&
  \begin{cases}
    e^{i\alpha_j}\sin\phi_j\prod_{k=1}^{j-1}\cos\phi_k&,\ j<N\\
    e^{i\alpha_N}\prod_{k=1}^{N-1}\cos\phi_k&,\ j=N
  \end{cases}\\
  r_{j,k}:=&\label{eq:R2}
  \begin{cases}
    0&,\ j<k<N-1\\
    e^{i\alpha_k}\cos\phi_k&,\ j=k<N-1\\
    -e^{i\alpha_j}\sin\phi_k\sin\phi_j\prod_{l=k+1}^{j-1}\cos\phi_l&,\ k<j\le N-1\\
    -e^{i\alpha_N}\sin\phi_k\prod_{l=k+1}^{N-1}\cos\phi_l&,\ k<N-1\ \wedge\ j=N
  \end{cases}\\
  r_{j,N-1}:=&\label{eq:R3}
  \begin{cases}
    0&,\ j<N-1\\
    e^{-i\neg_{N-1}}\cos\phi_{N-1}&,\ j=N-1\\
    -e^{-i\neg_N}\sin\phi_{N-1}&,\ j=N\\
  \end{cases}
\end{align}
where $\neg_j:=\sum_{k=1}^{j-1}\alpha_k+\sum_{k=j+1}^N\alpha_k$. Then, the vectors $R_{N,k}:=(r_{1,k},r_{2,k},\ldots,r_{N,k})^T$
are orthonormal and $s\in\{0,1\}$ can be chosen such that $R_N:=\begin{pmatrix}(-1)^sR_{N,1}&R_{N,2}&R_{N,3}&\cdots&R_{N,N}\end{pmatrix}\in SU(N)$. The parameters $\alpha_j\in[0,2\pi)$ and $\phi_j\in\left[0,\frac\pi2\right]$ in the definition of $R_N$ are uniquely defined by $R_{N,N}\in\d B_{\cn[N]}\cong S^{2N-1}$ up to a set of measure zero ($r_{j,N}=0\ \Leftrightarrow\ \alpha_j$ non-unique). Hence, we can define
\begin{align}
  \Phi_N:\quad \d B_{\cn[N]}\times\bigtimes_{j=2}^{N-1}\d B_{\cn[j]}\to SU(N);\quad (R_{N,N},r)\mapsto R_Nj_N(\Phi_{N-1}(r))
\end{align}
everywhere up to a set of measure zero.

Given suitable quadrature rules $Q_j$ in $\d B_{\cn[j]}$ such that none of the points in $\bigtimes_{j=2}^N Q_j$ is in the null set that $\Phi_N$ is not defined on, we obtain a quadrature rule 
\begin{align}
  Q_{SU(N)}:=\Phi_N\left[\bigtimes_{j=2}^N Q_j\right]
\end{align}
in $SU(N)$. Furthermore, the Haar measure of $U(N)$ decomposes into $h_{U(N)}=h_{SU(N)}\times h_{U(1)}$ (cf., e.g., \cite[Exercise 2.1.7]{abbaspour-moskowitz}), i.e., choosing a quadrature rule $Q_{U(1)}$ yields a quadrature rule $Q_{U(N)}:=Q_{SU(N)}\times Q_{U(1)}$ on $U(N)$.

As we are interested in polynomially exact quadrature rules, we choose
\begin{align}
  Q_{U(1)}:=\left\{e^{\frac{2\pi i k}{t+1}};\ k\in\nn_{\le t+1}\right\}
\end{align}
on $U(1)$ with equal weights $\frac{1}{t+1}$ which integrates all polynomials up to degree $t$ exactly \cite[Example 5.14]{delsarte}. Furthermore, since the pull-back of polynomials in $SU(N)$ have similar symmetry properties to ``standard'' polynomials on spheres, we consider (randomized) fully symmetric quadrature rules as described in \cite{genz} for the quadrature rules $Q_j$ on $S^{2j-1}$.

\begin{remark*}
  It should be noted that presently we do not have a complete proof of polynomial exactness for the groups $U(N)$ and $SU(N)$ with general $N$. However, the application below provides numerical evidence for the tested groups and polynomial degrees.
\end{remark*}

\section{One dimensional lattice QCD}\label{sec:1dQCD}
Let us consider the Dirac operator for a quark of mass $m$ at chemical potential $\mu$ \cite{ravagli}
\begin{align}
  \begin{aligned}
    \Df(U)
    =
    \begin{pmatrix}
      m&\frac{e^\mu}{2} U_{1}&&&&\frac{e^{-\mu}}{2} U_{n}^*\\
      -\frac{e^{-\mu}}{2} U_{1}^*&m&\frac{e^\mu}{2} U_{2}&&&\\
      &-\frac{e^{-\mu}}{2} U_{2}^*&m&\frac{e^\mu}{2} U_{3}&&\\
      &&\ddots&\ddots&\ddots&\\
      &&&-\frac{e^{-\mu}}{2} U_{n-2}^*&m&\frac{e^{-\mu}}{2} U_{n-1}\\
      -\frac{e^\mu}{2} U_{n}&&&&-\frac{e^{-\mu}}{2} U_{n-1}^*&m
    \end{pmatrix}
  \end{aligned}
\end{align}
where all empty entries are zero and the corresponding one flavor partition function 
\begin{align}
  &Z(m,\mu,G,n)=\int_{G^n}\det\Df(U)\ dh_G^n(U)
\end{align}
where $G\in\{U(N),SU(N)\}$, $N\in \nn$,  and $h_G$ is the corresponding (normalized) Haar measure on $G$. (Note that the integrals $Z(m,\mu,G,n)$ can be evaluated analytically; \cite[Theorem 2.2]{jansen} or \cite{bilic-demeterfi} if $n\in 2\nn$.)

As an observable for the model, we will consider the chiral condensate
\begin{align}
  \chi(m,\mu,G,n)=\d_m\ln Z(m,\mu,G,n)=\frac{\d_mZ(m,\mu,G,n)}{Z(m,\mu,G,n)}=\frac{\int_{G^n}\d_m\det\Df(U)\ dh_G^n(U)}{\int_{G^n}\det\Df(U)\ dh_G^n(U)}.
\end{align}
We will furthermore choose the gauge $U_j=1$ except $U_n=U$ which yields
\begin{align}
  \det\Df=&\det\left(\prod_{j=1}^{n}\tilde m_j+2^{-n}e^{-n\mu}U^*+(-1)^{n}2^{-n}e^{n\mu}U\right)
\end{align}
where $\tilde m_1:=m$, $\fa j\in[2,n-1]\cap\nn:\ \tilde m_j:=m+\frac{1}{4\tilde m_{j-1}}$, and $\tilde m_n:=m+\frac{1}{4\tilde m_{n-1}}+\sum_{j=1}^{n-1}\frac{(-1)^{j+1}2^{-2j}}{\tilde m_j\prod_{k=1}^{j-1}\tilde m_k^2}$. In particular, this reduces the integration over $G^n$ to an integral over $G$.

\section{Numerical Results}\label{sec:num-res}

\begin{figure}
\centering
\begin{minipage}{.45\textwidth}
  \centering
  \includegraphics[width=.9\linewidth]{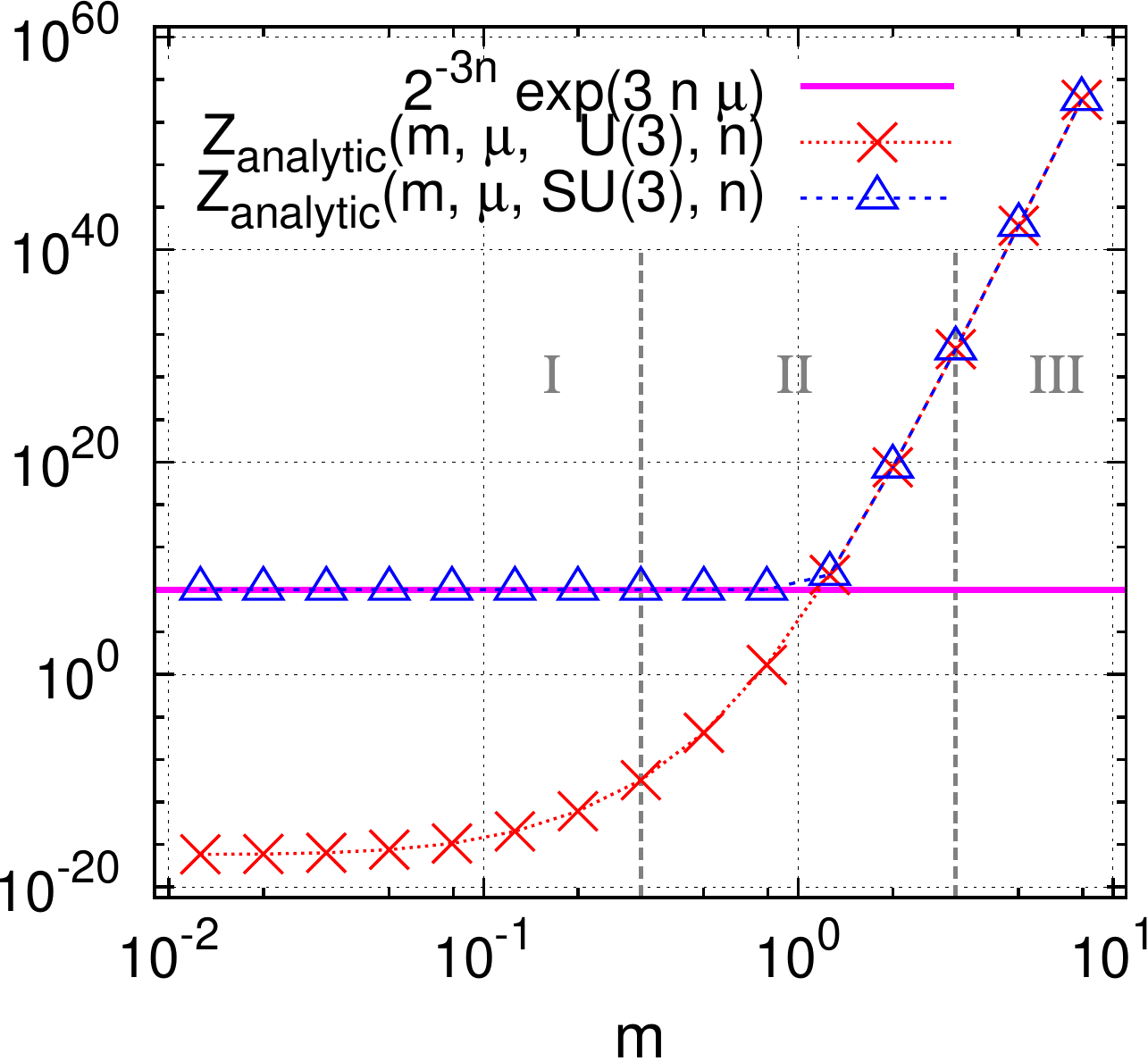}
  \captionof{figure}{\footnotesize Order of magnitude $2^{-3n}e^{3n\mu}$ of the point evaluation of the integrand compared to the value of $Z$ for $G\in\{U(3),SU(3)\}$.}
  \label{Fig2_val}
\end{minipage}\qquad %
\begin{minipage}{.45\textwidth}
  \centering
  \includegraphics[width=.7\linewidth]{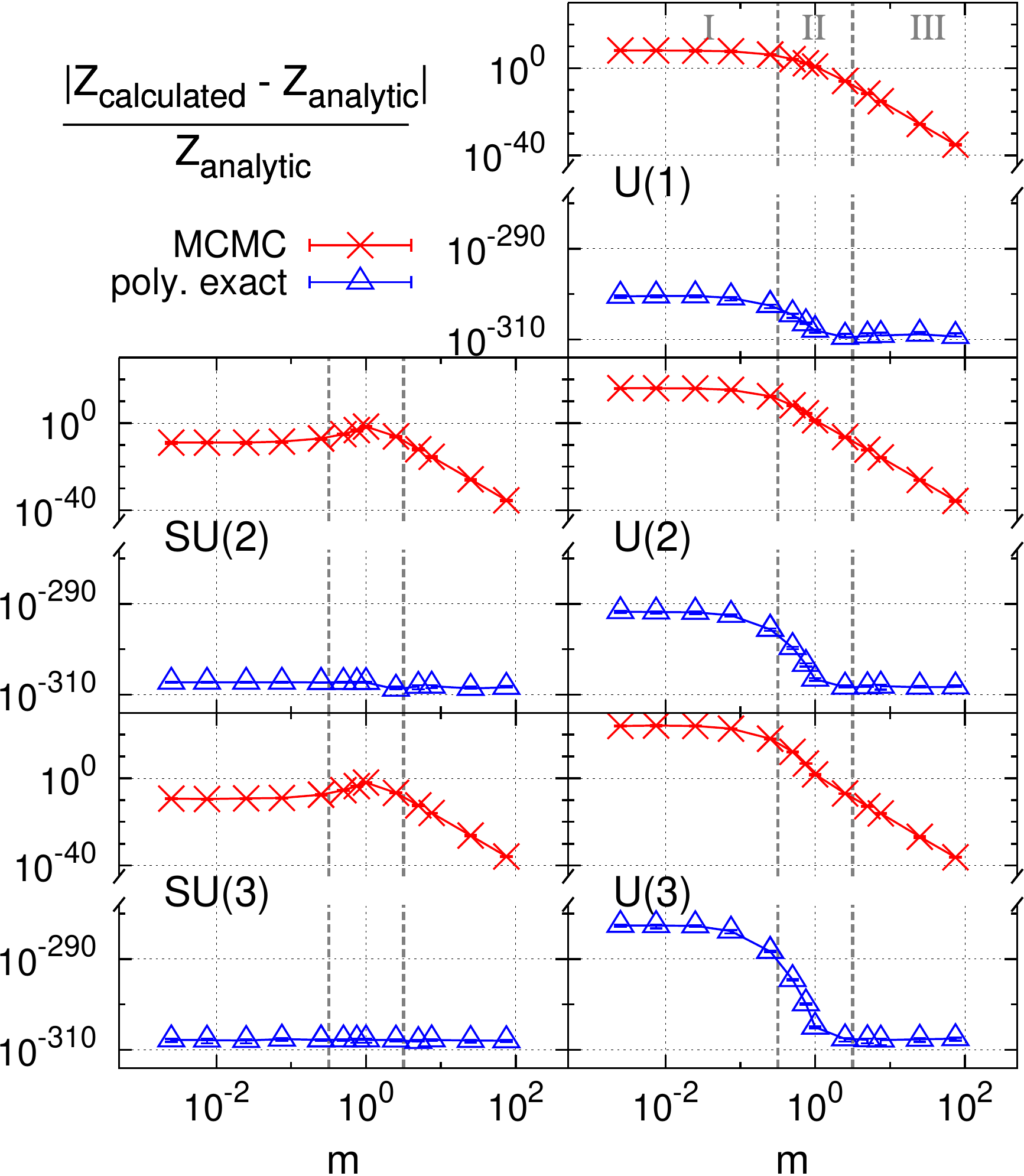}
  \captionof{figure}{\footnotesize Comparison of the relative error of $Z$ using MC-MC and polynomially exact quadrature. Computations were performed with $1024$bit extended floats ($\approx307$ digits precision).}
  \label{Fig3}
\end{minipage}
\end{figure}

In this section, we will compare the quadrature rules described in section~\ref{sec:quadrature} to Markov Chain Monte Carlo (MC-MC) using the same number of integration points. As a prelude to these comparisons, Figure~\ref{Fig2_val} shows the values of $Z(m,\mu=1.,G,n=20)$ with $G\in\{SU(3),U(3)\}$ over a range of $m$ values and compares them to $2^{-3n}e^{3n\mu}$ which is the order of magnitude of the non-constant term in $\det\Df$. We can identify three regions (I, II, and III).

For large values of $m$ (region III) the value of $Z$ is significantly larger than the order of magnitude of the critical terms in the point evaluation. Thus, in region III, we expect both methods (MC-MC and polynomially exact) to yield very small relative errors. 

For small values of $m$ (region I), on the other hand, $Z(m,\mu,U(3),n)$ is significantly lower than $2^{-3n}e^{3n\mu}$. The initial error will, therefore, be very large and MC-MC will produce very large relative errors. The polynomially exact quadrature should still produce machine error results, but since the point evaluation error is of the same order of magnitude as the MC-MC error (which is larger than $O(1)$), machine precision results still lose some precision. Since $Z(m,\mu,SU(3),n)-Z(m,\mu,U(3),n)\approx 2^{-3n}e^{3n\mu}$ we expect a regularizing effect on the relative error for $SU(3)$ in region~I, i.e., the relative error in the MC-MC case will not be as bad as it is in the $U(3)$ case.

In the transition region (region II) of intermediate $m$ values, we expect the MC-MC error in the $U(3)$ case to smoothly transition from machine error (region III) to some very large relative error (region I). The polynomially exact results should stay on machine error until the MC-MC error grows above $1$ and then smoothly transition to the machine precision results of region I. The $SU(3)$ case, however, could show a more interesting behavior since the constant term $Z(m,\mu,SU(3),n)-Z(m,\mu,U(3),n)\approx 2^{-3n}e^{3n\mu}$ and the critical term in the point evaluation are of the same order of magnitude. This can lead to cancelation effects (particularly for MC-MC) and, thus, a relative error that is larger than the relative error in both regions I and III. 

In Figure~\ref{Fig3}, we have plotted the relative error of $Z(m,\mu=1.,G,n=20)$ over a range of $m$ values and for all $G\in\{U(1),U(2),SU(2),U(3),SU(3)\}$. We have performed the computations with $1024$bit extended floating point arithmetic since the values of $Z(m\ll1.,\mu=1.,U(N),n=20)$ are too low for double precision (as can be seen in Figure~\ref{Fig2_val}). In particular, we observe that the polynomially exact method proposed above operates on machine precision throughout the entire range of $m$ values while MC-MC yields the expected results in each of the regions I, II, and III.

\begin{figure}
\centering
\begin{minipage}{.45\textwidth}
  \centering
  \includegraphics[width=.95\linewidth]{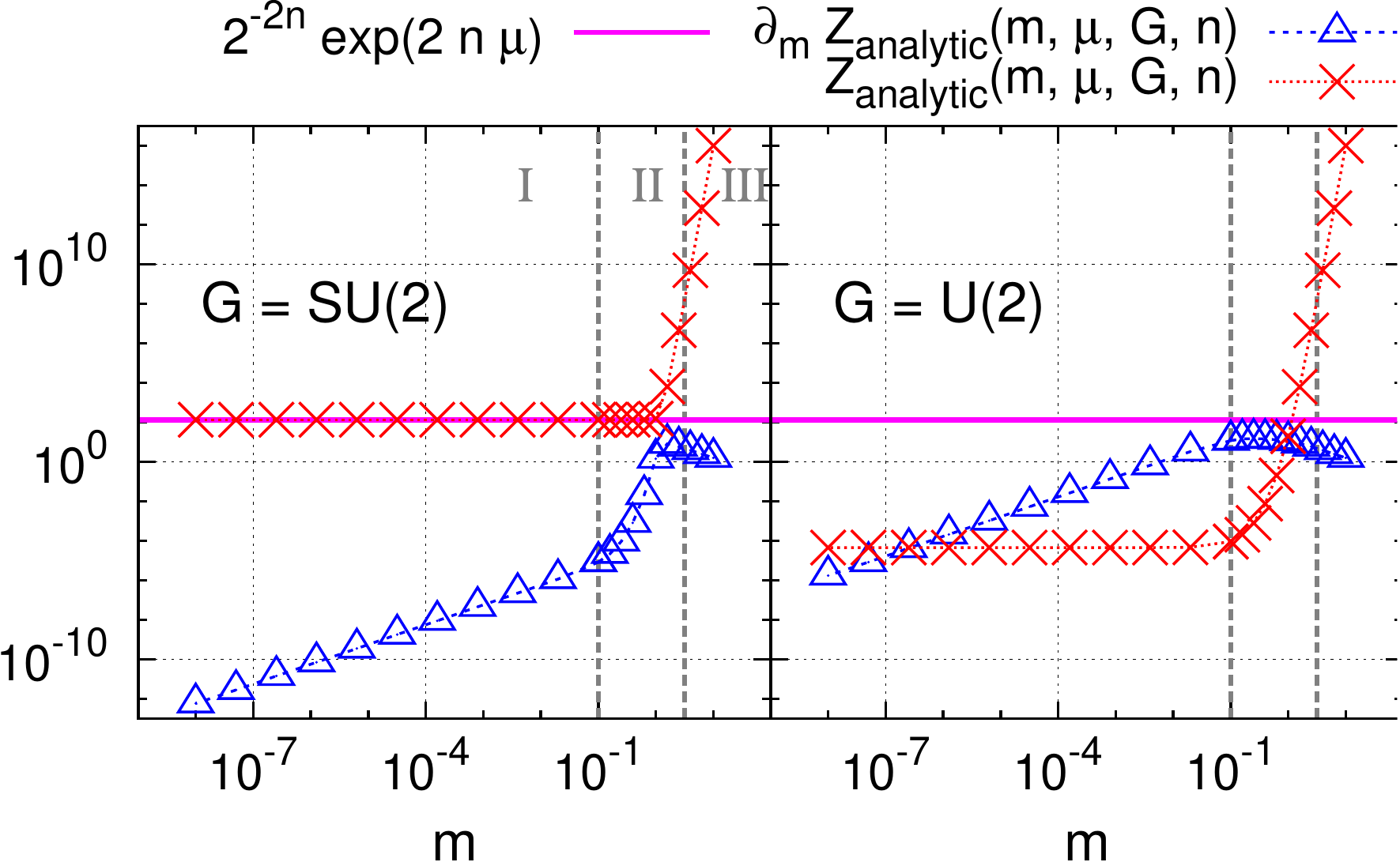}
  \captionof{figure}{\footnotesize Order of magnitude $2^{-2n}e^{2n\mu}$ of the point evaluation of the integrand compared to the values of $Z$ and $\d_mZ$ for $G\in\{U(2),SU(2)\}$. }
  \label{Fig4_val}
\end{minipage}\qquad %
\begin{minipage}{.45\textwidth}
  \centering
  \includegraphics[width=.7\linewidth]{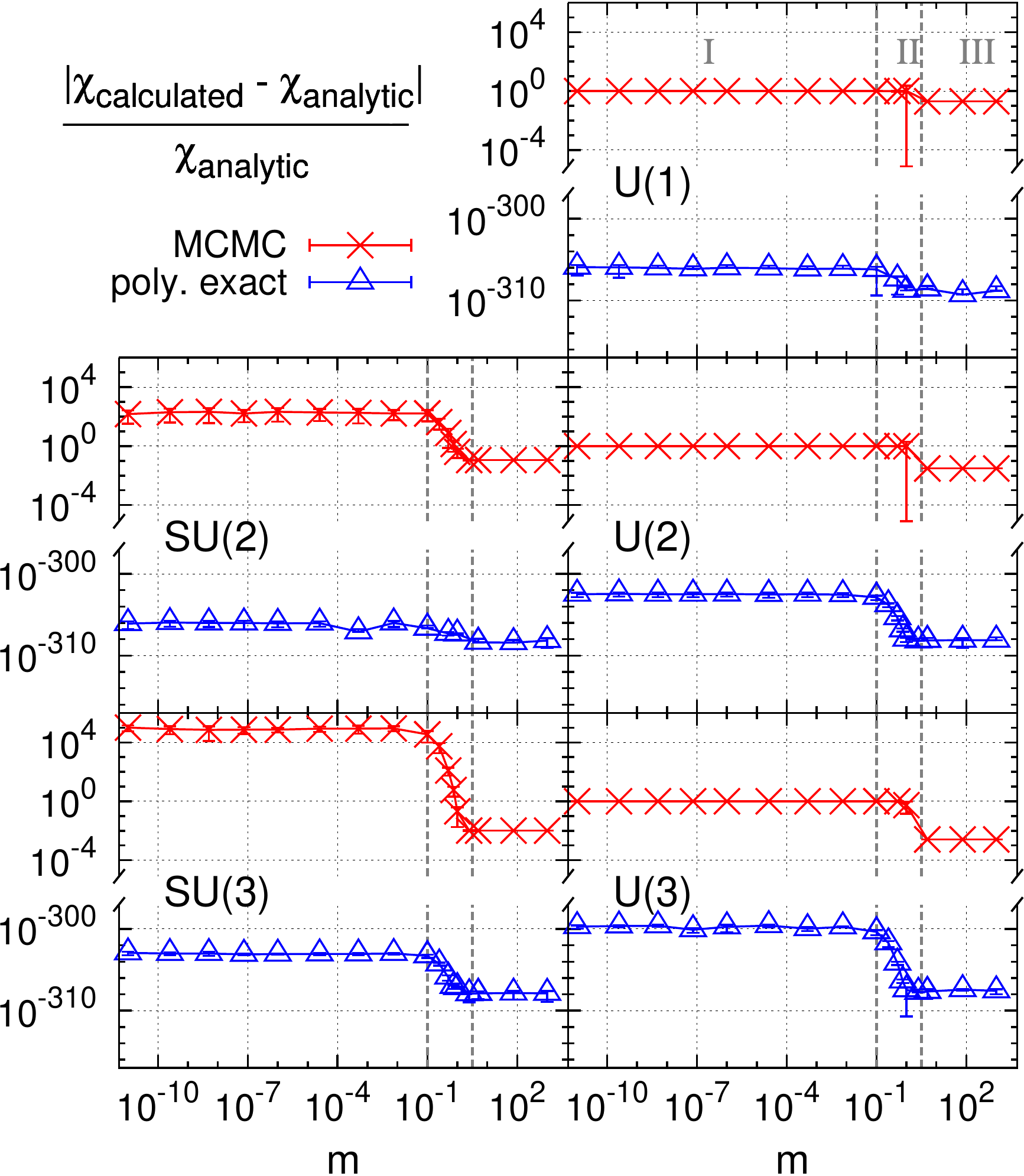}
  \captionof{figure}{\footnotesize Comparison of the relative error of $\chi$ using MC-MC and polynomially exact quadrature. Computations were performed with $1024$bit extended floats ($\approx307$ digits precision).}
  \label{Fig4}
\end{minipage}
\end{figure}

Considering the chiral condensate
\begin{align}
  \chi(m,\mu,G,n)=\frac{\d_mZ(m,\mu,G,n)}{Z(m,\mu,G,n)},
\end{align}
Figure~\ref{Fig4_val} shows the values of $Z(m,\mu=1.,G,n=8)$ and $\d_mZ(m,\mu=1.,G,n=8)$ over a large range of $m$ values and with $G\in\{U(2),SU(2)\}$, as well as, the order of magnitude $2^{-2n}e^{2n\mu}$ of the point evaluation. Adding to the complications of computing the denominator $Z$ for small $m$, the numerator $\d_mZ$ is difficult to compute, as well. Thus, we expect MC-MC results similar to the MC-MC results computing partition function $Z(m,\mu,U(N),n)$; this is precisely what we observe in Figure~\ref{Fig4}. In particular, we note that the MC-MC relative error is $O(1)$ for small $m$, i.e., no statistically significant results for the chiral condensate can be obtained using standard MC-MC methods. In contrast, the polynomially exact results are on machine precision as expected.

\section{Conclusion}\label{sec:conclusion}
We have developed new integration rules for the groups $U(N)$ and $SU(N)$, provided numerical verification that these rules are polynomially exact for $N\le3$, and compared them to Markov Chain Monte Carlo using the example of $1$-dimensional QCD with a chemical potential in which a sign problem appears with Monte Carlo for certain parameter values (region~I; $\prod_{j=1}^n\tilde m_j< 2^{-Nn}e^{Nn\mu}$). These computations have shown that, even in parameter ranges with the most severe sign problem, the chiral condensate can be computed to arbitrary precision using the newly proposed method. Standard Markov Chain Monte Carlo methods, on the other hand, exhibit large $O(1)$ relative errors, i.e., not giving any statistically significant results. We even used $1024$bit extended precision for these comparisons and obtained machine precision results with the new method. Furthermore, the newly constructed quadrature rules yield an error reduction by many orders of magnitude compared to Monte Carlo in regions without the sign problem, as well. Hence, we conclude that our polynomially exact method completely avoids the sign problem.

The fact that these new integration rules overcome the sign problem and reduce the error by orders of magnitude in the $1$-dimensional QCD is very promising and a notable result in its own right. However, this benchmark should be regarded as a toy model as it is necessary to demonstrate applicability of the method to higher dimensions. We are, thus, presently considering the Schwinger model as a $2$-dimensional quantum field theory. 

\acknowledgments
The authors wish to express their gratitude to Prof. Andreas Griewank for inspiring comments and conversations, which helped to develop the work in this article. H.L. and J.V. acknowledge financial support by the DFG-funded projects JA 674/6-1 and GR 705/13.

\end{document}